# SME Gender-Related Innovation: A Non-Numerical Trend Analysis Using Positive, Zero, and Negative Quantities

Nina Bočková[1], Barbora Volná[2*], Mirko Dohnal[3]

[1] Prague University of Economics and Business, Faculty of Business Administration, Department of Entrepreneurship, W. Churchilla 4, 130 67 Praha 3, Czech Republic, e-mail: nina.bockova@vse.cz
[2] Mathematical Institute in Opava, Silesian University in Opava, Na Rybníčku 1, 746 01 Opava, Czech Republic, e-mail: Barbora.Volna@math.slu.cz
[3] Brno University of Technology, Faculty of Business and Management, Kolejní 2906/4, Brno 612 00, Czech Republic, e-mail: dohnal@fbm.vutbr.cz
*Corresponding author

## Abstract

### *Purpose*

The primary objective of this paper is to examine the gender-related aspects of complex innovation processes in Small and Medium Enterprises (SMEs). SME innovations are unique, subjective, partially inconsistent, multidisciplinary, uncertain, and multi-dimensional. Moreover, the datasets and knowledge bases concerning Gender Aspects of SME Innovations (GASI) are often insufficient for the development of traditional formal (analytical/statistical) models.

This paper analyses the quality and quantity of data required to develop a sufficiently accurate GASI model. A ten-dimensional GASI model is studied using variables such as *Gender, Product Innovation, Process Innovation, and High-Risk Tolerance*.

### *Why this paper?*

Classical statistical analysis of GASI requires a high level of information intensity, which creates pressure to apply alternative formal tools. While these tools may not be as precise as statistical methods, they can incorporate vague, non-numerical information, such as: *SMEs managed by male entrepreneurs tend to be more innovative than those managed by female entrepreneurs*. Artificial intelligence algorithms, including fuzzy and rough sets, are widely used to mitigate the negative consequences of data scarcity. This paper identifies the best possible approach to quantifying GASI – through trend-based modelling.

### *Design/methodology/approach*

Artificial Intelligence (common sense reasoning) has developed models using just trends, i.e. *increasing, decreasing, and constant*. The trends are the least information intensive quantifiers and are therefore used to eliminate (minimise) problems related to GASI information shortages.

### *Findings*

A set of 13 ten-dimensional trend scenarios is identified, along with all possible transitions between them. This enables trend-based queries to be answered, such as: *Is it possible to increase exports, decrease product innovation and high-risk tolerance, while maintaining gender-related parameters constant?*

### *Original/value*

While extensive academic research has emphasized the importance of gender aspects in SME activities, innovation-related gender aspects have largely been overlooked. This study contributes to filling this gap.

### *What is new in the paper?*

Trend-based, low-information-intensity quantifiers (increasing, constant, decreasing) are employed for modelling GASI. Due to severe information shortages and the extreme heterogeneity of available data, traditional numerical quantifiers and statistical methods are prohibitively difficult to apply to GASI analysis.

### *Research limitations/implications*

The primary limitation of this study stems from the nature of trend-based models: no quantitative values are available. However, these limitations can be partially mitigated by integrating additional common-sense reasoning techniques within artificial intelligence methods.

### *Practical implications*

GASI trend models can be developed relatively easily and can incorporate a very high number of variables – ten or more. In contrast, quantitatively based GASI models, such as those using fuzzy sets, are not well suited for such multidimensional analyses.

**Keywords:** Gender, Innovation, Trend, Non-numerical quantification, Scenarios, Artificial intelligence.

**Introduction**

Women entrepreneurship has received substantial attention (see, e.g., Brush et al., 2020; Martínez-Rodríguez et al., 2022; Expósito et al., 2024). The literature analysing the impact of the entrepreneur's gender on business performance has significantly developed over the past decade (Kiefer et al., 2022; Lemma et al., 2022).

Women managers may face greater obstacles in implementing innovations compared to men, due to more limited access to financial resources, a lack of human capital, and weaker integration into innovation networks, among other factors (Vosta and Jalilvand, 2014).

Recent studies have shown that the entrepreneur's gender influences a firm's innovation behaviour, particularly in terms of the types of innovation implemented (Expósito et al., 2020; Expósito et al., 2023). Similarly, other studies confirm this relationship (see, e.g., Martínez-Rodríguez et al., 2022; Expósito et al., 2024). However, the role of the entrepreneur's gender in shaping how innovations affect business performance remains insufficiently explored (see, e.g., Cowling et al., 2020).

It is a well-established fact that different genders demonstrate substantially distinct attitudes toward management and business in general (see, e.g., Sarango-Lalangui et al., 2023; Joensuu-Salo et al., 2024). The studied GASI (Gender Aspects of SME Innovation) problem is highly heterogeneous, interdisciplinary, and context-specific, making its traditional statistical analysis prohibitively difficult (Medina-Vidal et al., 2025). Moreover, severe information shortages related to GASI create pressure to apply atypical, non-traditional analytical methods. The high information intensity required by classical statistical approaches drives computer experts to develop new formal tools – or upgrade existing ones – capable of handling such challenges (see, e.g., Dohnal and Doubravský, 2016). These tools may not be as precise as traditional statistical methods but are able to incorporate vague, non-numerical information items, such as the following (Expósito et al., 2024):

- SMEs managed by male entrepreneurs demonstrate higher innovativeness than those managed by female entrepreneurs.
- The presence of women in management positions positively influences corporate social responsibility.
- Innovativeness has a positive impact on both financial and operational business performance.

GASI-related forecasts represent a broad spectrum of complex tasks that are difficult to observe. These tasks are often unique and interdisciplinary in nature, involving fields such as macroeconomics, law, and engineering. GASI-related tasks can be analysed from various perspectives, including, for example, the political situation. In other words, there exists a wide range of different models.

Various tools are used to analyse GASI-related tasks, including statistical analysis, fuzzy or rough sets, genetic algorithms, and other methods of artificial intelligence. As a result, several types of quantitative simplifications are commonly applied – for example, linearization. These simplifications often lead to oversimplified models, which in turn produce results that are frequently inapplicable in practice (Dohnal and Doubravský, 2015).

This paper addresses GASI forecasting under conditions of severe information shortages. Such shortages are likely to cause significant issues when traditional statistical methods are employed.

The complexities of real-life GASI tasks make any formal description challenging. Sets of input information and knowledge items are extremely heterogeneous. The following list illustrates typical categories:

- Dominantly Subjective Information
    - Experience
    - Analogy
- Partially Subjective Information
    - Own observations or measurements
    - Commercially available observations
    - Literature sources
    - Verbal descriptions
- Dominantly Objective Information
    - Mathematical models (e.g., systems of differential equations):
        - Without numerical values of parameters
        - With known values of constants and parameters
    - Statistical models (e.g., polynomial functions based on the least squares algorithm):
        - Original data sets are available
        - No original data sets are available
        - Partial data set availability

AI technology has become so affordable that even small businesses can access and adopt it, depending on their needs and budgets (Nair and Gupta, 2021). Artificial intelligence methods have attracted significant attention in recent years (see, e.g., Falk et al., 2019). Compared with traditional statistical approaches, AI methods do not require strict assumptions about data distribution; they are also capable of handling large-scale datasets and capturing nonparametric and nonlinear relationships.

Common sense is essential for enhancing reasoning capabilities and minimizing problems caused by a lack of observations. In short, computers lack common sense.

**Data Collection**

SMEs contribute to over half of employment and GDP in most countries, regardless of their income levels. Moreover, fostering the development of SMEs can promote economic diversification and resilience (see, e.g., Al-Omoush et al., 2023). While it is not feasible to address every individual SME case (Saunders et al., 2019), the general economic rationale suggests that supporting more SMEs – where appropriate – can amplify the overall positive impact.

There are two fundamental activities associated with data collection techniques: first, the collection of datasets, and second, the analysis of that data (see, e.g., Bryman and Bell, 2011; Bryman, 2012). Any quantitative approach must inevitably rely on quantitative indicators, such as numerical or linguistic values (see, e.g., Van Raan, 2013; Devetak et al., 2010; Bell et al., 2022). Well-established techniques for collecting and analysing datasets are described in Tashakkori et al. (2020).

Forecasting and decision-making related to GASI are often based on models of unique or atypical companies. As a result, conventional statistical methods – typically relying, directly or indirectly, on the law of large numbers – are difficult or even impossible to apply. This implies that knowledge items with varying levels of subjectivity must be taken into account in order to develop the most appropriate model for the unique task under investigation. In particular, numerous observations of bankruptcy cases would be required; however, such data are generally unavailable.

This is why information non-intensive formal tools are being used increasingly often – see, e.g., fuzzy and/or rough sets (Pavláková Dočekalová and Kocmanová, 2016; Bočková et al., 2012).

Common sense is essential for enhancing reasoning capabilities and minimizing problems caused by limited observations. Computers inherently lack common sense. The formalization of common sense has attracted scholarly attention for quite some time (see, e.g., McCarthy, 1959; Lenat and Guha, 1990; Nakova et al., 2009).

A very vague and general statement – rooted in medieval or traditional interpretations of human common-sense reasoning – such as *"No tree grows to heaven"* or *"Nothing can grow beyond all limits"* can nevertheless be incorporated into trend-based models, for example, as:

$$A \text{ Quantitatively Unknown Upper Limit exists.} \tag{1}$$

There are several types of quantifiers, with numbers being the most frequently used (see, e.g., Atkinson et al. 2012; Vitanov and Vitanov, 2016).

**Trend Quantities**

Different quantifiers vary in their information intensity. The least information-intensive are trends – *decreasing, constant*, and *increasing*. If a trend cannot be quantified or predicted, then nothing can be measured or observed. Nevertheless, an optimal decision can still be made based on the trend alone (Wright and Goodwin, 2009).

The complete list of trend quantifiers is provided in Table 1 (see, e.g., Dohnal, 2016).

| Symbols: | + | 0 | – | * |
|---|---|---|---|---|
| Values: | Positive | Zero | Negative | Anything |
| Derivations: | Increasing | Constant | Decreasing | Any direction |

Table 1 – Trend Quantifiers Represented by Qualitative Values

**GASI Knowledge**

GASI knowledge is represented by a highly heterogeneous network of various items. These items differ in accuracy, may be partially inconsistent with one another, and originate from diverse domains such as economics, engineering, sociology, and politics. The following remarks illustrate key characteristics of GASI knowledge items:

- During the decision-making process, it is reasonable for experts to use linguistic variables to express their (semi-)subjective opinions.
- Modern data analysis techniques, partially based on artificial intelligence methods, do not replace GASI tasks performed by humans. Many companies invest time, effort, and resources into artificial intelligence, yet do not experience the anticipated benefits.

Statistical studies are based on the law of large numbers, which implies that a large number of GASI observations is required. As a result, unique and complex tasks can rarely be properly analysed using traditional statistical methods.

Various subfields of artificial intelligence can be applied to solve GASI (sub)tasks, including, for example, Naive Physics (Dohnal, 1988), ant colony optimization (Aliyu et al., 2021), fuzzy pooling (Dohnal and Kocmanová, 2016), fuzzy upgrade techniques (Chiarini, 2021), and knowledge-based hesitant fuzzy information.

**Trend Models**

As we mentioned before, in trend-based modelling we do not use quantitative values of model variables and parameters; we use only three qualitative values: +, – and 0. From this point of view, the trend model is represented by the set of possible qualitative values of model variables $X_1, X_2, \ldots, X_n$ and their first and second time derivatives, $DX_i$ and $DDX_i$ for the i-th variable.

Therefore, the *trend model* is given by the set of triplets

$$(X_i, DX_i, DDX_i), \qquad (2)$$

where $i = 1, 2, \ldots, n$. We say that model (2) *is solved* when all scenarios of possible qualitative values of triplets $(X_i, DX_i, DDX_i)$ are obtained for each variable $X_i$. Thus, the scenario set is given by

$$S = \{((X_1, DX_1, DDX_1), (X_2, DX_2, DDX_2), \ldots, (X_n, DX_n, DDX_n))_j\}, j = 1, 2, \ldots, m. \qquad (3)$$

The scenario set (3) is *n-dimensional* and contains *m* scenarios.

Moreover, the third time derivative, $DDDX_i$, can be used. However, such values are rarely known when GASI tasks are studied. Therefore, they are not considered in this paper.

Additional information that can be incorporated into the n-dimensional trend model consists of *w* pairwise relations between model variables $X_1, X_2, \ldots, X_n$ denoted by $P_k$, $k = 1, 2, \ldots w$:

$$\{P_1(X_{i1}, X_{j1}), P_2(X_{i2}, X_{j2}), \ldots, P_w(X_{iw}, X_{jw})\}. \qquad (4)$$

We can describe all pairwise relations $P(X,Y)$ for positive values of the variables $X$ and $Y$ using Table 2. The first column contains labels. The second column provides a verbal description of the function $Y(X)$. The third column presents a description in the form of a triplet of qualitative signs (+, –, or 0), representing the function value and its first and second derivatives, respectively. This triplet characterizes the qualitative behaviour of the function. The fourth column contains sketches of the graphs of these functions. These pairwise relations represent qualitative trend descriptions, meaning that no quantitative or numerical information is assumed to be known.

| Label | Description | Triplet | Graph Sketch | Label | Description | Triplet | Graph Sketch |
|---|---|---|---|---|---|---|---|
| CXI X Y | Convex increasing | + + + | 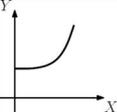 | CXD X Y | Convex decreasing | + − + | 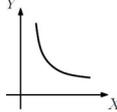 |
| LNI X Y | Linear increasing | + + 0 | 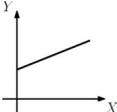 | LND X Y | Linear decreasing | + − 0 | 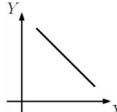 |
| CVI X Y | Concave increasing | + + − | 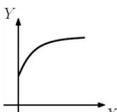 | CVD X Y | Concave decreasing | + − − | 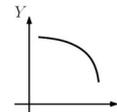 |

Table 2 – Representative Pairwise Relations Describing Trends

For example, the relation represented by the triplet (+ + +) indicates that the relation is increasing (the first derivative $dY/dX$ is positive), and the increase is becoming more and more rapid (the second derivative $d^2Y/dX^2$ is positive), i.e., a progressive increase. The positive sign of the function value implies that Y > 0 when X = 0. Similarly, the relation represented by (+ + −) stands for a degressive increase, where the relation is still increasing, but the rate of increase is slowing down. This corresponds to the previously discussed vague expression: *"No tree grows to heaven."*

Unfortunately, the second derivative of an economic quantity is often unknown or unavailable. In such cases, trend proportionalities based on first derivatives are considered instead. Let **DP** denote a direct qualitative proportionality and **IP** an indirect qualitative proportionality, where *X* and *Y* are model variables. Then:

DP X Y:  *If X is increasing, then Y is increasing. If X is decreasing, then Y is decreasing.*

IP X Y:  *If X is increasing, then Y is decreasing. If X is decreasing, then Y is increasing.* (5)

Another tool used in trend modelling is the so-called *trend correlation matrix*. Classical correlation matrices are frequently used in data analysis (see, e.g., Frigessi et al., 2011; Shevlyakov and Oja, 2016). It is possible to construct a trend model based on first-order trend derivatives, provided that a deterministic correlation matrix $C \in R^{n \times n}$ is available. The matrix **C** can be transformed into a trend model using the following rules:

$$\text{If } c_{i,j} > 0 \text{ then, DP } X_i X_j,$$

$$\text{If } c_{i,j} < 0 \text{ then, IP } X_i X_j. \tag{6}$$

The trend model generated from a numerical correlation matrix is *nearly always* inconsistent. The interpretation defined in (6) may produce a trend model that has no feasible solution – for example, a statement such as *"$X_i$ is increasing and decreasing at the same time."* This inconsistency arises from the statistical nature of the correlation matrix **C**, which may lead to incorrect or contradictory trend statements. Therefore, model (6) must be modified, and certain trend statements may need to be removed.

An obvious and simple removal algorithm, denoted *R*, can be described as follows:

*R: Remove the correlation coefficient $c_a$ with the smallest absolute value, and test the resulting matrix $C^R$. If the trend solution corresponding to the modified matrix represents the steady-state scenario:*

$$\begin{array}{ccccc} X_1 & X_2 & X_3 & \dots & X_n \\ +0* & +0* & +0* & \dots & +0* \end{array}$$

*then repeat this heuristic R.* (7)

Moreover, the generated model (6) must be confronted with all available additional sources of information. In practice, however, human common sense based on experience is often the only such source of information or knowledge.

**Confrontation of Trend GASI Models**

The accuracy of GASI models is often rather low. It is therefore highly desirable to compare the results of multiple models developed independently by different forecasters or decision-makers.

A team of *r* forecasters is involved

$$F_1, F_2, \dots F_r \tag{8}$$

It is usually not possible to achieve a full consensus within a team of *r* forecasters. For this reason, each forecaster develops their own n-dimensional model:

$$M_1^{(n)}, M_2^{(n)}, \dots M_r^{(n)} \tag{9}$$

Each of these models is solved, resulting in a set of trend-based n-dimensional scenarios:

$$S_1^{(n)}, S_2^{(n)}, \dots S_r^{(n)} \tag{10}$$

The *core* and *envelope* sets of scenarios (10) are defined as follows (see, e.g., Dohnal and Doubravský, 2015):

$$S_{COR}^{(n)} = S_1^{(n)} \cap S_2^{(n)}, ... \cap S_r^{(n)}$$
$$S_{ENV}^{(n)} = S_1^{(n)} \cup S_2^{(n)}, ... \cup S_r^{(n)} \quad (11)$$

The *core* set eliminates all atypical scenarios, while the *envelope* set covers all possible scenarios generated by all decision-makers. It is obvious that $S_{ENV}^{(n)}$ is a superset of $S_{COR}^{(n)}$:

$$S_{ENV}^{(n)} \supseteq S_{COR}^{(n)} \quad (12)$$

**Transitional Graphs**

The set **S** of m scenarios (see Equation (2)) is not the only output of trend analysis. It is also possible to generate the set

$$T = \{t_1, t_2, ..., t_r\} \quad (13)$$

of time transitions among the scenarios in **S**. In GASI trend modelling, we consider only scenarios with positive function values, as only positive values of the modelled economic quantity are meaningful in this case. Table 3 provides a set of all possible one-dimensional transitions involving a positive function value – that is, transitions where the first position in the corresponding triplet is "+". The one-dimensional case corresponds to n=1 in Equation (2).

| No. | From | To / Alternatives | | | | | | |
|---|---|---|---|---|---|---|---|---|
| 1 | + + + | + + 0 | | | | | | |
| 2 | + + 0 | + + + | + + − | | | | | |
| 3 | + + − | + + 0 | + 0 − | + 0 0 | | | | |
| 4 | + 0 + | + + + | | | | | | |
| 5 | + 0 0 | + + + | + − − | | | | | |
| 6 | + 0 − | + − − | | | | | | |
| 7 | + − + | + − 0 | + 0 + | + 0 0 | 0 − + | 0 0 + | 0 0 0 | 0 − 0 |
| 8 | + − 0 | + − + | + − − | 0 − 0 | | | | |
| 9 | + − − | + − 0 | 0 − − | 0 − 0 | | | | |

Table 3 – A List of All One-Dimensional Transitions Between Triplets With a Positive Function Value

For example, the third row of Table 3 illustrates that the triplet (+ + −) can transition into the triplet (+ + 0), or (+ 0 −), or (+ 0 0), respectively. The principle behind such transitions lies in the time behaviour of the economic quantity. This quantity changes over time continuously and smoothly, without any "jumps", "kinks" (i.e., sharp, non-smooth bends), or "steps". In other words, the corresponding time function, which represents the trend of this economic quantity, is continuous, smooth, or at least continuously differentiable. An example of such a "smooth" economic function *X* with positive function values is shown in Figure 1. This figure illustrates the simple, one-dimensional trend oscillation of function X over time and depicts possible transitions between the considered triplets (see Table 3).

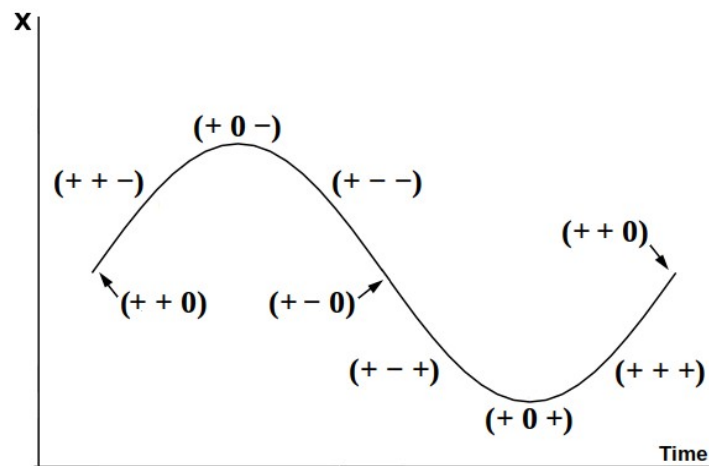
Figure 1 – Illustration of Trend Oscillations and Triplet Transitions

The graphical representation of the set **T** of time transitions among the scenarios in **S** is the so-called *transitional graph*. A transitional graph H is a directed graph in which the nodes represent the scenarios from the set **S**, and the directed arcs correspond to the transitions from the set **T**.

Any forecast in trend modelling represents a selection of a path through the corresponding transitional graph. In line with the previous discussion about individual forecasters, it is possible that each of them selects a different path through the transitional graph.

**Case Study**

Any trend analysis is inherently a combinatorial problem. There is a strong correlation between the number of scenarios and the number of variables. If the number of scenarios exceeds 50, the corresponding transitional graph becomes highly complex, and its interpretation is practically impossible. Cases involving several hundred scenarios are not uncommon.

Variables with a significant impact on GASI forecasting were therefore carefully selected to reflect the heterogeneous nature of the tasks under study. To support this process, a discussion was organized with a small team of experts and PhD students.

GASI trend models represent a broad spectrum of different innovation tasks. The following trend model is based on a trend-oriented reinterpretation of subjectively acquired GASI knowledge items. The integration of published heterogeneous knowledge sources has been a key contribution to the development of the model (see, e.g., McQuillan, 2023; Piosik, 2023; Perrin, 2023; Yuanjian and Wenxiang, 2020; Zhao et al., 2021; Welch et al., 2008).

A significant advantage of trend models is that there is no need to define variables with precision. It is important to keep in mind that only trend-based quantifications are required. For example, if the variable HRT (High-risk tolerance) is increasing, it is irrelevant whether its exact definition is provided, as long as it is understood based on common sense. The set of model variables is stated in Table 4. Thus, these variables are considered trend variables, and their exact descriptions are not required. Units such as USD, EUR, or mass in kilograms are irrelevant, as the quantifiers are not numerical values.

| Variable | Abbreviation |
|---|---|
| Gender | GEN |
| Product innovation | PRI |
| Process innovation | PRO |
| High-risk tolerance | HRT |
| Age of entrepreneur | AGI |
| University education | UNI |
| Small business | SMA |
| Medium business | MED |
| Age of business | AGE |
| Export | EXP |

Table 4 – The Set of Model Variables

Several trend interpretations of correlation matrices (6) were applied, and a sequence of removals (7) led to the first version of the GASI trend model, which is presented in Table 5. The first column contains the row number of the model. The second column lists the model row, where the first abbreviation indicates the mutual relationship (see Equation (6) and Table 2) between two model variables (see Table 4).

| No. | Model Row |
|---|---|
| 1 | CVI GEN AGI |
| 2 | DP GEN INI |
| 3 | DP GEN EXP |
| 4 | IP GEN PRI |
| 5 | DP HRT PRI |
| 6 | IP AGE PRI |
| 7 | IP AGE PRO |
| 8 | DP UNI HRT |
| 9 | DP MED PRO |
| 10 | DP SMA EXP |

Table 5 – First GASI Trend Model

Table 6 presents the set of scenarios derived from the GASI trend model (see Table 5). We can see that 13 scenarios are obtained, and scenario No. 7 represents the steady state of the model.

| No. | GEN | AGE | SMA | EXP | PRO | PRI | HRT | UNI | MED | AGI |
|---|---|---|---|---|---|---|---|---|---|---|
| 1 | + + + | + + + | + + + | + + + | + − − | + − − | + − − | + − − | + − − | + + + |
| 2 | + + + | + + + | + + + | + + + | + − − | + − − | + − − | + − − | + − − | + + 0 |
| 3 | + + + | + + + | + + + | + + + | + − − | + − − | + − − | + − − | + − − | + + − |
| 4 | + + 0 | + + 0 | + + 0 | + + 0 | + − 0 | + − 0 | + − 0 | + − 0 | + − 0 | + + − |
| 5 | + + − | + + − | + + − | + + − | + − + | + − + | + − + | + − + | + − + | + + − |
| 6 | + 0 + | + 0 + | + 0 + | + 0 + | + 0 − | + 0 − | + 0 − | + 0 − | + 0 − | + 0 + |
| 7 | + 0 0 | + 0 0 | + 0 0 | + 0 0 | + 0 0 | + 0 0 | + 0 0 | + 0 0 | + 0 0 | + 0 0 |
| 8 | + 0 − | + 0 − | + 0 − | + 0 − | + 0 + | + 0 + | + 0 + | + 0 + | + 0 + | + 0 − |
| 9 | + − + | + − + | + − + | + − + | + + − | + + − | + + − | + + − | + + − | + − + |
| 10 | + − + | + − + | + − + | + − + | + + − | + + − | + + − | + + − | + + − | + − 0 |
| 11 | + − + | + − + | + − + | + − + | + + − | + + − | + + − | + + − | + + − | + − − |
| 12 | + − 0 | + − 0 | + − 0 | + − 0 | + + 0 | + + 0 | + + 0 | + + 0 | + + 0 | + − − |
| 13 | + − − | + − − | + − − | + − − | + + + | + + + | + + + | + + + | + + + | + − − |

Table 6 – Scenarios of the First GASI Trend Model

It is straightforward to observe that the following three subsets of model variables result in identical scenarios (see Table 6) when processed by the trend model:

- GEN, AGE, SMA, EXP;
- PRO, PRI, HRT, UNI, MED;
- AGI.

This means, for example, that if the studied company is small (i.e., the variable SMA, see Table 4) and exports (EXP) need to be increased, then it is desirable to increase GEN and AGE, according to the above grouping of model variables.

The transitional graph based on the set of scenarios (see Table 6) is shown in Figure 2.

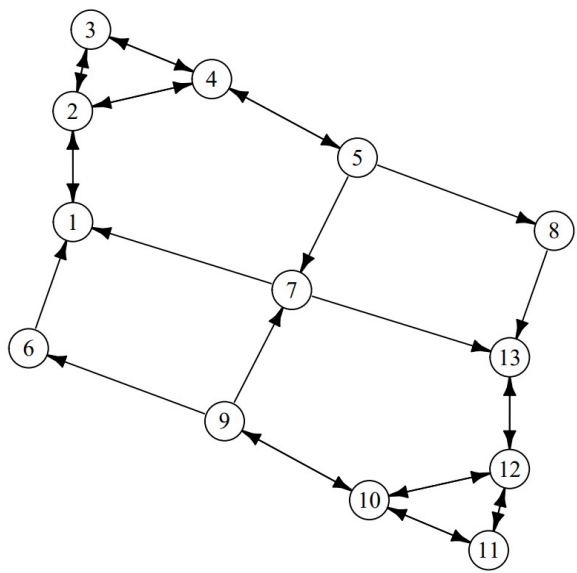

Figure 2 – Transitional Graph of the First GASI Trend Model

As illustrated in the transitional graph (see Figure 2), the model is stable. In this context, stability means the existence of a so-called *stabilisation loop* – i.e., a directed cycle that includes the steady state (in this model, scenario No. 7). Examples of such loops include:

$$7 \to 1 \to 2 \to 3 \to 4 \to 5 \to 7, \text{ or } 7 \to 13 \to 12 \to 11 \to 10 \to 9 \to 7.$$

Moreover, scenarios 6 and 8 can also be part of a stabilisation loop. For example:
$$7 \to 13 \to 12 \to 10 \to 9 \to 6 \to 1 \to 2 \to 4 \to 5 \to 7,$$
$$\text{or } 7 \to 1 \to 2 \to 4 \to 5 \to 8 \to 13 \to 12 \to 10 \to 9 \to 7.$$

From this, it follows that any economic setting of the model variables (model scenarios, see Table 6) allows for a return to the steady state (scenario No. 7), which represents an equilibrium. Even scenarios involving "extreme" cases – such as a progressive increase (+ + +) in some model variable, as seen in scenarios No. 1, 2, 3, and 13 – can lead to the steady state over time in this model. This corresponds to our initial assumption: *"No tree grows to heaven"* (see Equation (1)).

However, the team of cooperating experts does not believe that it is possible to move from scenario No. 7 (the steady state) to scenario No. 1 of GEN (+ + +) – i.e., a rapid increase – in a single step (see Table 6 and Figure 2). Therefore, several modifications of the GASI trend model were necessary in order to develop a version accepted by the majority of experts (see Table 7). As in the previous table, the rows of the modified model provide the mutual relationship (see Equation (6) and Table 2) between two model variables (see Table 4). The main modifications concern the specification of the mutual relationships in model rows 6 and 7 – where a very general indirect qualitative proportionality was replaced by a convex and a concave decrease, respectively.

| No. | Model Row |
|---|---|
| 1 | CVI GEN AGI |
| 2 | DP GEN INI |
| 3 | DP GEN EXP |
| 4 | IP GEN PRI |
| 5 | DP HRT PRI |
| 6 | CXD AGE PRI |
| 7 | CVD AGE PRO |
| 8 | DP UNI HRT |
| 9 | DP MED PRO |
| 10 | DP SMA EXP |

Table 7 – Expert-Modified GASI Trend Model

Table 8 provides the set of scenarios derived from the expert-modified GASI trend model (see Table 7). There are 21 scenarios, with scenario No. 11 representing the steady state of the model.

| No. | GEN | AGE | SMA | EXP | PRO | PRI | HRT | UNI | MED | AGI |
|---|---|---|---|---|---|---|---|---|---|---|
| 1  | + + + | + + + | + + + | + + + | + − − | + − − | + − − | + − − | + − − | + + + |
| 2  | + + + | + + + | + + + | + + + | + − − | + − − | + − − | + − − | + − − | + + 0 |
| 3  | + + + | + + + | + + + | + + + | + − − | + − − | + − − | + − − | + − − | + + − |
| 4  | + + 0 | + + + | + + 0 | + + 0 | + − − | + − 0 | + − 0 | + − 0 | + − − | + + − |
| 5  | + + − | + + + | + + − | + + − | + − − | + − + | + − + | + − + | + − − | + + − |
| 6  | + + − | + + 0 | + + − | + + − | + − − | + − + | + − + | + − + | + − − | + + − |
| 7  | + + − | + + − | + + − | + + − | + − + | + − + | + − + | + − + | + − + | + + − |
| 8  | + + − | + + − | + + − | + + − | + − 0 | + − + | + − + | + − + | + − 0 | + + − |
| 9  | + + − | + + − | + + − | + + − | + − − | + − + | + − + | + − + | + − − | + + − |
| 10 | + 0 + | + 0 + | + 0 + | + 0 + | + 0 − | + 0 − | + 0 − | + 0 − | + 0 − | + 0 + |
| 11 | + 0 0 | + 0 0 | + 0 0 | + 0 0 | + 0 0 | + 0 0 | + 0 0 | + 0 0 | + 0 0 | + 0 0 |
| 12 | + 0 − | + 0 − | + 0 − | + 0 − | + 0 + | + 0 + | + 0 + | + 0 + | + 0 + | + 0 − |
| 13 | + − + | + − + | + − + | + − + | + + − | + + − | + + − | + + − | + + − | + − + |
| 14 | + − + | + − + | + − + | + − + | + + − | + + − | + + − | + + − | + + − | + − 0 |
| 15 | + − + | + − + | + − + | + − + | + + − | + + − | + + − | + + − | + + − | + − − |
| 16 | + − 0 | + − + | + − 0 | + − 0 | + + − | + + 0 | + + 0 | + + 0 | + + − | + − − |
| 17 | + − − | + − + | + − − | + − − | + + − | + + + | + + + | + + + | + + − | + − − |
| 18 | + − − | + − 0 | + − − | + − − | + + − | + + + | + + + | + + + | + + − | + − − |
| 19 | + − − | + − − | + − − | + − − | + + + | + + + | + + + | + + + | + + + | + − − |
| 20 | + − − | + − − | + − − | + − − | + + 0 | + + + | + + + | + + + | + + 0 | + − − |
| 21 | + − − | + − − | + − − | + − − | + + − | + + + | + + + | + + + | + + − | + − − |

Table 8 – Scenarios of the Expert-Modified GASI Trend Model

It can be observed that the expert-modified GASI trend model generates different groupings of model variables that result in identical scenarios (see Table 8), compared to those in the first GASI trend model (see Table 6):

- GEN, SMA, EXP;
- PRI, HRT, UNI;
- PRO, MED;
- AGE;
- AGI.

In contrast to the first model, the expert-modified GASI trend model does not indicate that GEN shares identical scenario behaviour with AGE, PRI, HRT, UNI, PRO, MED, or AGI. These variables now belong to different scenario groups (see Table 8), and thus no similar policy or adjustment recommendations involving GEN and these variables can be derived.

The transitional graph corresponding to the expert-modified GASI trend model, based on the set of scenarios (see Table 8), is shown in Figure 3.

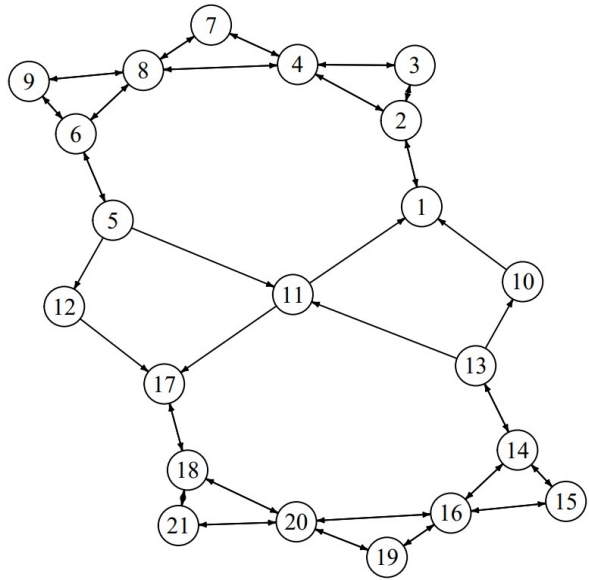

Figure 3 – Transitional Graph of the Expert-Modified GASI Trend Model

The expert-modified GASI trend model confirms the possibility of transitioning from the steady state (scenario No. 11) to a progressive increase of GEN (+ + +) through a sequence of scenarios: 11 → 1 → 2 → 3. The GEN triplet (+ + +) remains unchanged throughout the path 1 → 2 → 3, addressing the experts' concerns from the original model and supporting the plausibility of this development trajectory.

Similarly to the first version of the GASI trend model, we can observe several stabilisation loops in the expert-modified GASI trend model. For example, see the directed cycles:

11 → 1 → 2 → 4 → 8 → 6 → 5 → 11 or 11 → 17 → 18 → 20 → 16 → 14 → 13 → 11.

Furthermore, it is evident that all scenarios can be part of some stabilisation loop. For instance, scenarios 9, 12, and 19 are included in the loop:

11 → 1 → 2 → 4 → 8 → 9 → 6 → 5 → 12 → 17 → 18 → 20 → 19 → 16 → 14 → 13 → 11,

(see Figure 3). This confirms that, as in the first model, each scenario can eventually lead back to the steady state.

**Conclusion**

Currently, techniques used for analysing complex gender-related innovation processes are primarily analytical and/or statistical in nature. However, these precise mathematical tools do not always contribute as much as expected to a full understanding of the systems under study. It is therefore not paradoxical that information non-intensive methods of analysis often achieve more accurate and realistic results, particularly when the system being modelled is highly complex or poorly understood.

The main advantages of the proposed trend-based method are as follows:

- No numerical values of constants or parameters are required, and the set of trend-based solutions forms a superset of all meaningful solutions,
- A complete list of all feasible forecasts is obtained,

- Unsteady-state behaviours of trend-driven innovation processes can be identified,
- Results are easy to interpret without requiring knowledge of advanced mathematical tools,
- The method can serve as a starting point, with quantitative features incorporated in subsequent stages.

We presented two versions of the GASI trend model: the initial version and an expert-modified version. The first model was criticised by experts for allowing a direct transition from the steady state to a scenario with a rapidly increasing value of the "gender" variable in a single step. In response to this objection, the expert-modified version was developed. Interestingly, this model still preserves that trajectory, but now as part of a longer, continuous path involving the same progressive increase in "gender".

Additionally, the expert-modified model is stable, exhibiting multiple stabilisation loops. Since each scenario can be part of a stabilisation loop, this result aligns with the initial vague assumption that *"no tree grows to heaven."*

The integration of expert insights into the modelling process proved essential for aligning the trend model with real-world plausibility and decision-making logic. The proposed trend modelling framework is flexible and can be applied beyond gender-related innovation processes, especially in situations where expert knowledge is available, but precise data are lacking. This work demonstrates how qualitative trend models can bridge the gap between abstract system complexity and practically interpretable results.

**Acknowledgement**

This paper is supported by the Modelling and optimization of processes in the corporate sphere; Registration number FP-S-22-7977.

**References**

[1] Aliyu, M., Murali, M., Zhang, Z. J., Gital, A., Boukari, S., Huang, Y., Zahraddeen Yakubu, I. (2021), Management of Cloud Resources and Social Change in a Multi-Tier Environment: A Novel Finite Automata Using Ant Colony Optimization with Spanning Tree, *Technological Forecasting and Social Change* 166, 120591.

[2] Al-Omoush, K. S., Anderson, A., Ribeiro-Navarrete, B. (2023), The Impact of Absorptive Capacity, Corporate Social Innovation, and E-Business Proactiveness on SMEs' Survival, *Transformations in Business & Economics* 1 (58), 130–148.

[3] Atkinson, M. P., Kress, M., Szechtman, R. (2012), Carrots, Sticks and Fog during Insurgencies, Mathematical Social Sciences 64 (3), 203–13.

[4] Bell, E., Bryman, A., Harley, B. (2022), *Business research methods,* Oxford university press.

[5] Bočková, N., Brož, Z., Dohnal, M. (2012), Fuzzy model of relationship among economic performance, competitiveness and business ethics of small and medium-sized enterprises, Acta universitatis agriculturae et silviculturae Mendelianae Brunensis, LX (4), 71–78.

[6] Brush, C. G., Greene, P. G., Welter, F. (2020), The Diana project: A legacy for research on gender in entrepreneurship, *International Journal of Gender and Entrepreneurship* 12 (1), 7–25.

[7] Bryman, A. (2012), *Social Research Methods,* 4th ed., New York: Oxford University Press.

[8] Bryman, A., Bell, E. (2011), *Business Research Methods,* 3rd ed., Oxford University Press.

[9] Chiarini, A. (2021), Industry 4.0 Technologies in the Manufacturing Sector: Are We Sure They Are All Relevant for Environmental Performance? *Business Strategy and the Environment* 30 (7), 3194–3207.

[10] Cowling, M., Marlow, S., Liu, W. (2020), Gender and bank lending after the global financial crisis: Are women entrepreneurs safer bets? *Small Business Economics* 55 (4), 853–880.

[11] Devetak, I., Glažar, S. A., Vogrinc, J. (2010), The Role of Qualitative Research in Science Education, *Eurasia Journal of Mathematics, Science and Technology Education* 6 (1), 77–84.

[12] Dohnal, M. (1988), Naive Models as Active Expert System in Bioengineering and Chemical Engineering, *Collection of Czechoslovak Chemical Communications* 53 (7), 1476–1499.

[13] Dohnal, M. (2016), Complex Biofuels Related Scenarios Generated by Qualitative Reasoning under Severe Information Shortages: A Review, *Renewable and Sustainable Energy Reviews* 65, 676–84.

[14] Dohnal, M., Doubravský, K. (2015), Qualitative Upper and Lower Approximations of Complex Nonlinear Chaotic and Nonchaotic Models, *International Journal of Bifurcation and Chaos* 25 (13), 1550173.

[15] Dohnal, M., Doubravský, K. (2016), Equationless and Equation-Based Trend Models of Prohibitively Complex Technological and Related Forecasts, *Technological Forecasting and Social Change* 111, 297–304.

[16] Dohnal, M., Kocmanová, A. (2016), Qualitative Models of Complex Sustainability Systems Using Integrations of Equations and Equationless Knowledge Items Generated by Several Experts, *Ecological Indicators* 62, 201–211.

[17] Expósito, A., Sanchis-Llopis, J. A. (2020), The effects of innovation on the decisions of exporting and/or importing in SMEs: Empirical evidence in the case of Spain, *Small Business Economics* 55 (3), 813–829.

[18] Expósito, A., Sanchis-Llopis, A., Sanchis-Llopis, J. A. (2023), CEO gender and SMEs innovativeness: Evidence for Spanish businesses, *International Entrepreneurship and Management Journal* 19, 1017–1054.

[19] Expósito, A., Sanchis-Llopis, A., SanchisaLlopis, J. A. (2024), Entrepreneur's Gender and SMEs Performance: the Mediating Effect of Innovations, *Journal of the Knowledge Economy* 15, 11877–11911.

[20] Falk, T., Mai, D., Bensch, R. *et al.* (2019) U-Net: deep learning for cell counting, detection, and morphometry. *Nature Methods* 16, 67–70.

[21] Frigessi, A., Løland, A., Pievatolo, A., Ruggeri, F. (2011), Statistical Rehabilitation of Improper Correlation Matrices, *Quantitative Finance* 11 (7), 1081–1090.

[22] Joensuu-Salo, S., Kangas, E., Könönen, L., Koivuranta, A. (2024), Gendered Perspectives on Sustainable Entrepreneurship: A Study of Finnish SMEs, *Proceedings of the International Conference on Gender Research* 7, 167–74.

[23] Kiefer, K., Heileman, M., Pett, T. L. (2022). Does gender still matter? An examination of small business performance, *Small Business Economics* 58, 141–167.

[24] Lemma, T. T., Gwatidzo, T., Mlilo, M. (2022), Gender differences in business performance: Evidence from Kenya and South Africa, *Small Business Economics* 60 (2), 591–614.

[25] Lenat, D.B., Guha R. V. (1990), *Building Large Knowledge-Based Systems: Representation and Inference in the Cyc Project*, Addison-Wesley.

[26] Martínez-Rodríguez, I., Quintana-Rojo, C., Gento, P., Callejas-Albinana, F. E. (2022), Public policy recommendations for promoting female entrepreneurship in Europe, *International Entrepreneurship and Management Journal* 18, 1235–1262.

[27] McCarthy, J. (1959), *Programs with common sense*, Stanford University, Stanford, CA 94305.

[28] McQuillan, N. (2023), The role of entrepreneurial leadership in generating augmented SME growth an exploratory study, Doctoral Thesis, *Ulster University Business School*.

[29] Medina-Vidal, A., Alonso-Galicia, P. E., González-Mendoza, M., Ramírez-Montoya M. S. (2025), Financial inclusion of vulnerable sectors with a gender perspective: risk analysis model with artificial intelligence based on complex thinking, *Journal of Innovation and Entrepreneurship* 14:4.

[30] Nair, K., Gupta, R. (2021), Application of AI technology in modern digital marketing environment, *World Journal of Entrepreneurship, Management and Sustainable Development* 17 (3), 318–328.

[31] Nakova, E., Linnebank, F.E., Bredeweg, B., Salles, P., Uzunov, Y. (2009), The river Mesta case study: A qualitative model of dissolved oxygen in aquatic ecosystems, *Ecological Informatics* 4, (5-6), 339–357.

[32] Pavláková Dočekalová, M., Kocmanová, A. (2016), Composite indicator for measuring corporate sustainability, *Ecological Indicators* 61 (2), 612–623.

[33] Perrin, C. (2023), Gender, financial inclusion, and entrepreneurship, Economics and Finance. *Université de Strasbourg*, 2023. English.

[34] Piosik, J. (2023), Essays on Entrepreneurial Finance, Ph.D. thesis, *Copenhagen Business School*, PhD Series No. 43.2023.

[35] Sarango-Lalangui, P., Castillo-Vergara, M., Carrasco-Carvajal, O., Durendez, A. (2023), Impact of environmental sustainability on open innovation in SMEs: An empirical study considering the moderating effect of gender, *Heliyon* 9 (9), e20096.

[36] Saunders, M., Lewis, P., Thornhill, A. (2019). *Research methods for business students*, 8th ed., Pearson.

[37] Shevlyakov, G., L., Oja, H. (2016), *Robust Correlation: Theory and Applications*. John Wiley & Sons.

[38] Tashakkori, A., Johnson, R. B., Teddlie, C. (2020), *Foundations of mixed methods research: Integrating quantitative and qualitative approaches in the social and behavioral sciences*, Sage publications.

[39] Van raan, A. F. J. (2013), *Handbook of quantitative studies of science and technology*, 1st ed., Elsevier.

[40] Vitanov, N. K., Vitanov, K. N. (2016), Box Model of Migration Channels, *Mathematical Social Sciences* 80, 108–114.

[41] Vosta, L. N., Jalilvand, M. R. (2014), Examining the influence of social capital on rural women entrepreneurship: An empirical study in Iran, *World Journal of Entrepreneurship, Management and Sustainable Development* 10 (3), 209–227.

[42] Welch, C.L., Welch, D.E., Hewerdine, L. (2008), Gender and Export Behaviour: Evidence from Women-Owned Enterprises, *Journal of Business Ethics* 83, 113–126.

[43] Wright, G., and Goodwin, P. (2009), Decision Making and Planning under Low Levels of Predictability: Enhancing the Scenario Method, *International Journal of Forecasting* 25 (4), 813–25.

[44] Yuanjian, Q., Wenxiang, W., (2020), Research on the Relationship among Uncertainty Tolerance, Decision-Making Logic and Radical Innovation Performance, *Science & Technology Progress and Policy* 37 (02), 1–9.

[45] Zhao, H., O'Connor, G., Wu, J., Lumpkin, G.T. (2021), Age and entrepreneurial career success: A review and a meta-analysis, *Journal of Business Venturing* 36 (1), 106007.